\documentclass{ws-ijmpd}

\begin{document}

\markboth{V. Bosch-Ramon}
{Instructions for Typing Manuscripts (Paper's Title)}

%
\catchline{}{}{}{}{}
%

\title{Secondary emission behind the radio outflows in gamma-ray binaries?}

\author{Valent\'i Bosch-Ramon}

\address{Departament d'Astronomia i Meteorologia and 
Institut de Ci\`encies del Cosmos (ICC), Universitat de Barcelona (UB/IEEC), Mart\'{\i} i Franqu\`es 1,
08028, Barcelona, Spain\\
Max Planck Institut f\"ur Kernphysik, Saupfercheckweg 1, Heidelberg 69117, Germany
\\
vbosch@mpi-hd.mpg.de}

\maketitle

\begin{history}
\received{Day Month Year}
\revised{Day Month Year}
\comby{Managing Editor}
\end{history}

\begin{abstract}

Several binary systems consisting of a massive star and a compact object have been detected above 100~GeV in the Galaxy. In
most of these sources, gamma-rays show a modulation associated to the orbital motion, which means that the emitter should
not be too far from the bright primary star. This implies that gamma-ray absorption will be non negligible, and large amounts
of secondary 
electron-positron pairs will be created in the stellar surroundings. In this work, we show that the radio emission from
these pairs should be accounted for when interpreting the radio spectrum, variability, and morphology found in gamma-ray
binaries. Relevant features of the secondary 
radio emission are the relatively hard spectrum, the orbital motion of the radio peak
center, and the extended radio structure following a spiral-like trajectory. The impact of the stellar wind free-free
absorption should not be neglected. 

\end{abstract}


\section{Introduction}	

Five galactic sources of very high-energy (VHE) gamma-rays have been associated so far to binary systems consisting of a
compact object and a massive bright star: PSR~B1259$-$63\cite{aha05a}; LS~5039\cite{aha05b}; LS~I~+61~303\cite{albert06};
Cygnus~X-1\cite{albert07}; HESS~J0632$+$057\cite{hinton09}. Three of them show a modulation in gamma-rays associated to the
orbital motion (PSR~B1259$-$63\cite{aha05a}; LS~5039\cite{aha06}; LS~I~+61~303\cite{albert09}), which implies that the
gamma-ray emitter cannot be too far from the star. Since the spectrum of the star radiation peaks in the ultraviolet,
efficient absorption of VHE photons will take place through interaction with stellar photons, leading to the creation in the
stellar surroundings of secondary electron-positron pairs with energies $\ge 10$~GeV, which is about a half of pair creation
gamma-ray energy threshold. Once created, secondary pairs cool down trapped in the stellar wind due to slow diffusion and
radiate mainly via synchrotron and inverse Compton (IC) emission. Depending on the medium magnetic field, either synchrotron
or IC will dominate (for the latter, see Ref.~\refcite{pelliza09}). The gamma-ray binaries found to date emit also
non-thermal radio and X-ray radiation (see Ref.~\refcite{bosch09} and references therein). In principle, despite the primary
gamma-ray emitter could be also producing radiation at lower energies, the component generated by the secondary pairs in the
stellar wind could even dominate the whole non-thermal output of the source\cite{bosch08}. 

\section{General picture}

The multiwavelength radiation produced by secondary pairs created in the stellar environment has been studied in detail using
a semi-analytical approach in Ref.~\refcite{bosch08}. However, in that work the complex 3-dimensional structure of the
secondary trajectories was simplified anchoring the secondary pairs to the stellar wind\footnote{This is reasonable as long as
the particle mean free path is much smaller than the system size.}, with the latter being taken as spherically symmetric,
with constant velocity, and wind and magnetic field energy density in the wind $\propto 1/r^2$ ($r$: distance to the
star). In addition, the magnetic field was considered fully irregular. For simplicity, the adiabatic cooling of the secondary
pairs was switched off, which is a reasonable approximation for relatively compact systems and at binary spatial scales, but
not for the extended radio emission produced well outside the binary system. Finally, free-free absorption in the stellar
wind was not considered. This approach can give a reasonable estimate of the radio fluxes, not as good for the spectrum, and
provides little spatial information. 

In this work, we are interested in the spectrum, variability, and the structure of the emitting secondary pairs at radio
wavelengths. For an appropriate study of the radio emission, a detailed follow-up of the secondary spatial and energy
evolution in the system region, accounting for the conditions of the stellar wind, is required. Given the difficulties of
treating the magnetic field structure and adiabatic cooling in the wind in a semi-analytical approach, we have performed
numerical calculations in which secondary pairs are injected and tracked using a Monte-Carlo code. First, we calculate the
injected spatial and energy distribution of pairs continuously produced by pair creation through the interaction of the
primary gamma-rays and the stellar photons. In the Monte-Carlo simulation, the primary gamma-ray energy distribution is
$dN_{\gamma}(E)/dE\propto E^{-1}$ for convenience. Thus, we must weight each particle according to the real primary gamma-ray
distribution to obtain physical results. Once the secondary pair distribution in space and energy is computed, the radio
synchrotron emission can be calculated taking into account the observer direction and the local magnetic field geometry. The
emission for different areas of the sky plane can be obtained, which allows to produce synthetic radio maps convolving the
fluxes in these different sky areas with a circular Gaussian of FWHM=1~milliarcsecond (mas). The result would be the image as
observed by a radio interferometer with angular resolution of 1~mas. 

\section{Monte-Carlo simulations of secondary pairs in gamma-ray binaries}

Once secondary pairs are produced in the stellar surroundings, since they are embedded in the magnetized stellar wind, they
start to spiral moving along the magnetic field lines and diffuse to some extent, provided the likely presence of some
irregular field. Particles cool down through ionization/coulombian interactions with wind atoms and ions, relativistic
Bremsstrahlung, adiabatic cooling as long as particles are confined in the wind, and synchrotron and IC radiation.  The
energy evolution of the secondary pairs is computed following the particle trajectories with suitable time steps. The steps
have been chosen such that the particle energy does not change significantly due to cooling in each of them. In the
calculations presented here the particle energy cannot decrease more than a 20\% in each step. In addition, since particles 
move in the stellar wind, which is neither homogeneous in density nor in magnetic field, the distance covered by secondary
pairs in each time step should not be larger than the characteristic variation length of the density and magnetic field, and
the same applies in relation to the stellar photon field. 

The wind density depends on $r$ as $n(r)=\dot{M}_{\rm w}/4\,\pi\,r^2\,v_{\rm w}(r)$,
where $\dot{M}_{\rm w}$, $v_{\rm w}(r)\approx v_{\rm w\infty}\,(1-r_*/r)$, $r_*$ and $v_{\rm w\infty}$ are
the star mass-loss rate, the radial wind velocity, the stellar radius, and the wind velocity at infinity, respectively.
The wind magnetic field can be described as
${\bf B_{\rm w}}(r,\phi)=(B_{\rm r},B_\phi)$, where $B_r\approx B_*(r_*/r)^2$, $B_\phi\approx B_*(v_{\rm w\phi}/v_{\rm w\infty})(r_*/r)$, $B_*$, and $v_{\rm w\phi}$
are the radial and toroidal magnetic field
components, the stellar surface magnetic field, and the initial toroidal component of the 
wind velocity, respectively. We neglect here the region within
the Alfven radius ($r_{\rm A}$), in which $B_{\rm w}$ is dipolar, 
since the $r$-values relevant here are in general $>r_{\rm A}\sim r_*$. See Ref.~\refcite{usov92} for a model of
the wind and its magnetic field.

Depending on the relation between $r$ and the mean free path of the particles, the trajectories are computed using either the
magnetic adiabatic invariant or diffusion, adopting an anisotropic coefficient
$D_{\parallel/\perp}=\lambda_{\parallel/\perp}\,c/3$, in which the particle mean free paths have been taken as
$\lambda_{\parallel}=\eta\,r_{\rm g}=\eta\,E/q_{\rm e}\,|B_{\rm w}|$ and $\lambda_\perp=r_{\rm g}/\eta$, where $\eta$ relates
the irregular and total magnetic field components as $\delta B=\eta^{-1/2}\,|B_{\rm w}|$. The symbols $\parallel$ and $\perp$
refer to the parallel and perpendicular directions to the magnetic field line. 

\section{Results}

We have applied the Monte-Carlo code sketched in the previous section to the particular case of the gamma-ray binary 
LS~5039. LS~5039 is an X-ray
binary that consists of a O6.5 main sequence star and a compact object of unclear nature (neutron star or black hole; see 
Ref.~\refcite{casares05}). The system semi-major axis is $a=2.2\times 10^{12}$~cm, with eccentricity $e=0.33$ and orbital
period $P=3.9$~days.  The phase 0.0 is taken at the periastron passage, and phases 0.05 and 0.67 are the superior and the
inferior conjunction of the compact object, respectively. The inclination angle is not known and could be in a range between
$i\approx 15^\circ-75^\circ$. The stellar mass-loss rate has been estimated
in $\sim 3\times 10^{-7}\,M_{\odot}/$yr. In this work, we have used the ephemeris, orbital parameters, system-observer
geometry, star and stellar wind properties, and inclination angle provided in Refs.~\refcite{casares05,aragona09}. 
We have fixed $i$ to $45^\circ$. 
The
parameters $B_*$ and $\eta$ have been fixed to reasonable values, $200$~G (see Ref~\refcite{bosch08b} for a discussion 
on this) and 100 (i.e. $\delta B=0.1\,|B_{\rm
w}|$), respectively. The gamma-ray distribution injected in the system, which leads to the pair creation, has been chosen
such that the source should show absorbed fluxes and photon indices roughly consistent with the observed ones, i.e. $\sim
10^{-11}$~erg~s$^{-1}$ (at 3~kpc) and $\sim 2.5$, respectively. To zeroth order, the location of the emitter has been taken
equal to that of the compact object. Note however that in LS~5035 this is unlikely true, as noted in Ref.~\refcite{bosch08b},
but this location choice has mainly impact close to the superior conjunction of the compact object. 

\begin{figure}[pb]
\centerline{\psfig{file=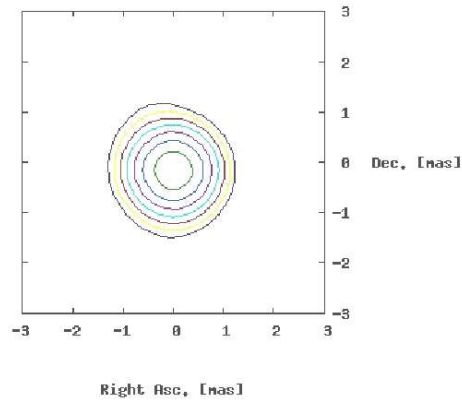,width=6.5cm}}
\vspace*{8pt}
\caption{Image of the fluxes per beam at 5~GHz for LS~5039, as seen by the observer using a radio VLBI interferometer 
with FWHM=1~mas. The orbital phase is 0.75, 
slightly after the inferior conjunction of the compact object. The system parameters adopted are from
Aragona et al. (2009), as well as the orientation in the sky plane (see fig.~6). The inclination has been 
fixed to $45^\circ$, which is an intermediate value within the possible range given by Casares et al. (2005). 
The contour levels 
start from 0.8~mJy/beam, the assumed telescope noise level, and grow $\times 2$ inwards.\label{f1}}
\end{figure}

\begin{figure}[pb]
\centerline{\psfig{file=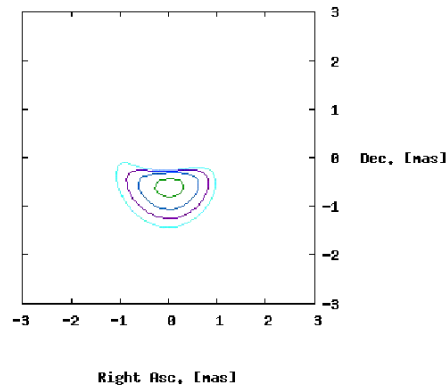,width=6.5cm}}
\vspace*{8pt}
\caption{Residual image after subtracting a pure gaussian to the image shown in Fig.~\ref{f1}, which would correspond to a
pure point-like source with the flux and location of the {\it real} one. 
Note that the residuals are 4~$\sigma$ above the noise
leve. The rest is as in Fig.~\ref{f1}.\label{f2}}
\end{figure}

\begin{figure}[pb]
\centerline{\psfig{file=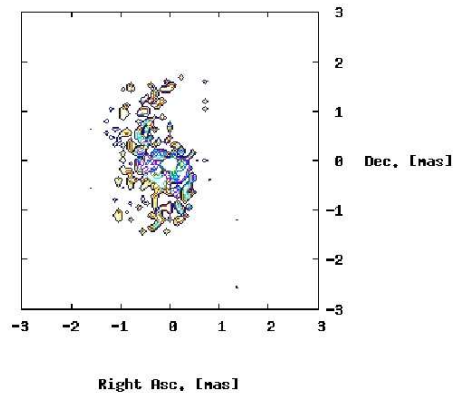,width=6.5cm}}
\vspace*{8pt}
\caption{Map of the fluxes per beam at 5~GHz projected in the observer plane for LS~5039. This image has not been 
convolved with a Gaussian. The pixel size is 
$3\,10^{12}\times 3\,10^{12}$~cm$^2$. Note that a spiral-like structure is hinted by the emission morphology (see 
Fig.~\ref{f4}), 
although the number of particles of the Monte-Carlo simulation does not allow
a smoother aspect for this image. The most of the emission comes from a region of $1\times 1$~mas$^2$, 
although the 
{\it spiral} is broad enough to yield the residuals shown in Fig.~\ref{f2}. The rest is as in Fig.~\ref{f1}.
\label{f3}}
\end{figure}

\begin{figure}[pb]
\centerline{\psfig{file=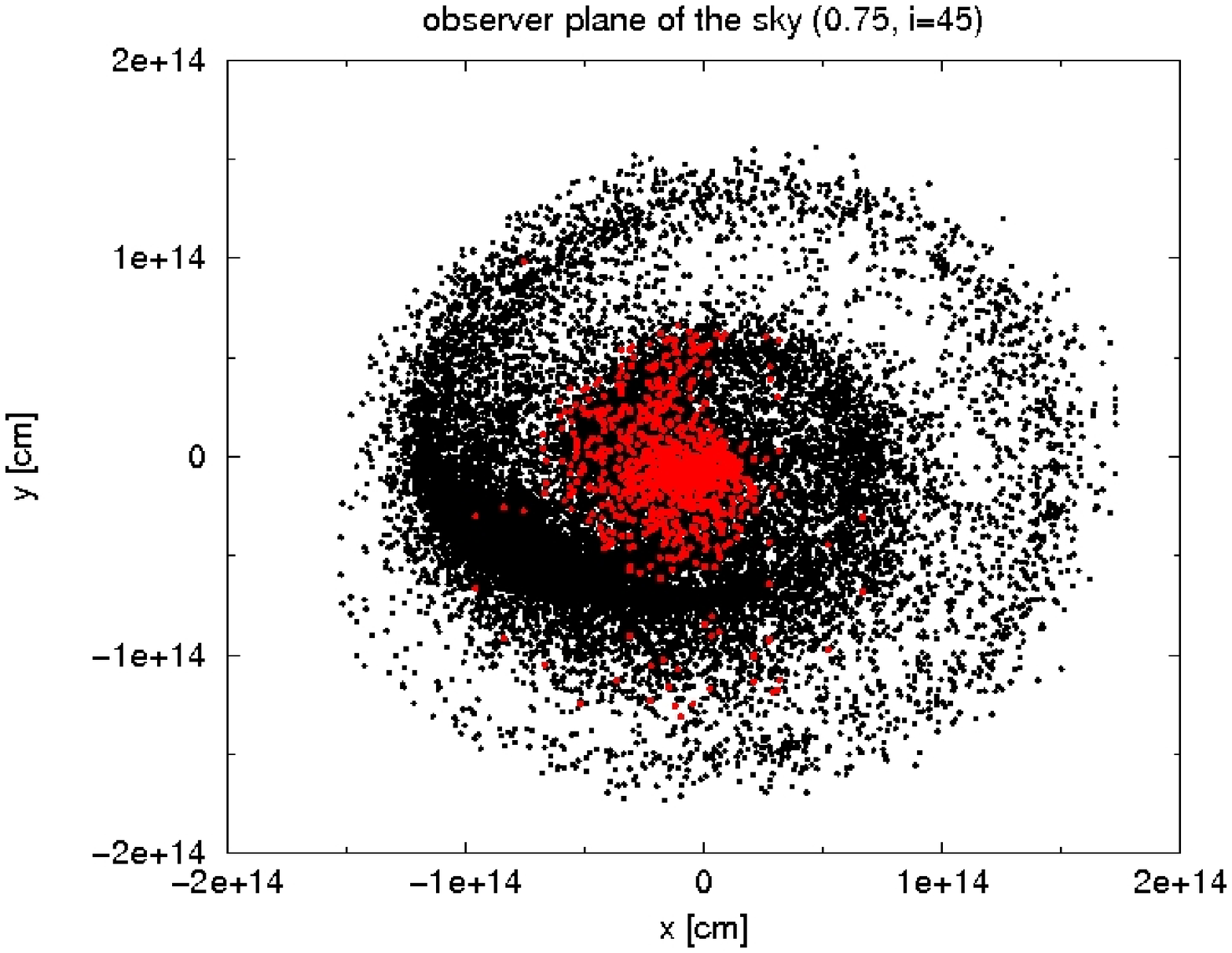,width=6.5cm}}
\vspace*{8pt}
\caption{Image of the secondary spatial distribution at phase 0.75 projected in the plane of the sky of the observer. 
This result has been obtained after running the code for two orbital periods ($\approx 7.8$~days).
In red, the electrons producing 5~GHz radio emission are shown. 
The radio images presented in Figs~\ref{f1}, \ref{f2} and \ref{f3} correspond to the red electron population.
The orbital parameters, inclination, and 
sky orientation are taken as in
Fig.~\ref{f1}. 
Note that the shown scales are about one hundred times larger than the binary system size.
\label{f4}}
\end{figure}

The results of the calculations are specific fluxes at 5~GHz of about 10~mJy adopting a relatively weak ionization wind
fraction of $X_{\rm i}=0.03$ to compute the wind free-free absorption. We recall that the distance to the star is at least
several $r_*$, and $X_{\rm i}$ is hard to estimate there. The spectra are close to flat, and even may be inverted for large
$X_{\rm i}$ values.  The morphological results are presented in Figs.~\ref{f1}, \ref{f2}, \ref{f3}, and \ref{f4} for the
phase 0.75. The emitter shape changes orientation along the orbit as long as the gamma-ray emitter-star spatial relation
changes,  and the peak of the radio emission moves about half mas along the orbit. A proper description of each figure can be
found in the caption. For a source as compact as LS~5039, the secondary radio emission is marginally extended under the
assumptions of our model, but it already shows hints of spiral-like structure when looking at the residuals. For more
extended sources, it would be possible to resolve this radiation. A moderate improvement in angular resolution 
would actually allow us to probe the
secondary emitter in detail.  However, we note that this emission is very sensitive to stellar wind (a)symmetries, the
magnetic field and the wind ionization level.  We conclude that the radio emission from secondary may be contributing to the
resolved core of LS~5039 found using VLBI techniques\cite{ribo08}, and should not be neglected when interpreting the radio
emission in gamma-ray binaries in general.

\section*{Acknowledgments}
V.B-R. acknowledges support by the Ministerio de Educaci\'on y Ciencia (Spain) 
under grant AYA 2007-68034-C03-01. 
V.B-R. wants to thank the Insituto Argentino de Astronom\'ia, and the Facultad de Ciencias Astron\'omicas y 
Geof\'isicas de la Universidad de La Plata, for their kind hospitality.


\end{document}